\documentstyle[12pt,a4,epsf]{article}
\makeatletter
\@addtoreset{equation}{section}
\makeatother

\newcommand{\VEV}[1]{\left\langle #1 \right\rangle}

\newcommand{\bequ}{\begin{equation}}
\newcommand{\eequ}{\end{equation}}
\newcommand{\beqn}{\begin{eqnarray}}
\newcommand{\eeqn}{\end{eqnarray}}

\begin{document}
\begin{titlepage}

\begin{flushright}
hep-ph/0211034\\
KUNS-1811\\

\today
\end{flushright}

\vspace{4ex}

\begin{center}
{\large \bf
Grand Unification with Anomalous U(1) Symmetry
\footnote
{Talk presented at SUSY2002 held at DESY, Humburg(Germany),
Jun. 17-23 2002.
}
}

\vspace{6ex}

\renewcommand{\thefootnote}{\alph{footnote}}
Nobuhiro Maekawa\footnote
{e-mail: maekawa@gauge.scphys.kyoto-u.ac.jp
}

\vspace{4ex}
{\it Department of Physics, Kyoto University, Kyoto 606-8502, Japan}
\end{center}

\renewcommand{\thefootnote}{\arabic{footnote}}
\setcounter{footnote}{0}
\vspace{6ex}

%--------------------<<   abstract   >>--------------------
\begin{abstract}
In this talk, we introduce a new scenario of grand unified theory
(GUT) with anomalous $U(1)_A$ gauge symmetry.
Since generic interactions (including non-renormalizable interactions)
are introduced, once we fix the symmetry of the theory, we can define
the theory without $O(1)$ coefficients.
The number of parameters to determine the symmetry is just 11 or 12
integer parameters which are anomalous $U(1)_A$ charges(integer)
for introduced multiplets.
It is surprising that the GUT scenario can explain
doublet-triplet splitting, quark and lepton
masses and mixing angles. In neutrino sector, the scenario
realizes LMA solution for solar neutrino problem and large 
$U_{e3}=O(0.1)$. Moreover, the scenario predicts
that the unification scale becomes smaller than the usual GUT
scale, namely proton decay via dimension 6 operators may be
seen in future experiments.
This talk is based on the papers\cite{maekawa,maekawa2,maekawa3,BM,MY}.

\end{abstract}

\end{titlepage}

%--------------------<<   section    >>--------------------
\section{Introduction}
Grand unified theory (GUT)
\cite{georgi} realizes two kinds of unification.
It unifies 3 gauge interactions 
$G_{SM}=SU(3)_C\times SU(2)_L\times U(1)_Y$ in the standard model 
into 1 gauge interaction like $SU(5)$, $SO(10)$ or $E_6$.
It is regarded as the signal for the unification of gauge interaction
that three gauge couplings meet at a scale 
$\Lambda_G\sim 2\times 10^{16}$ GeV in the minimal supersymmetric (SUSY)
standard model (MSSM). Moreover, GUT gives a unification
of quark and leptons in a fewer multiplets.
In the standard model, 6 multiplets ($Q$, $U^c$, $D^c$, $L$, $E^c$ and $N^c$)
are required for one family quark and lepton, while in the $SU(5)$ GUT, 
3 multiplets (${\bf 10}=Q+U^c+E^c$, ${\bf \bar 5}=D^c+L$, ${\bf 1}=N^c$)
for one family. If we adopt $SO(10)$ GUT, just one multiplet
(${\bf 16}={\bf 10}+{\bf \bar 5}+{\bf 1}$) is needed for one family.
However, introducing ${\bf 10}$ of $SO(10)$ naturally avoids unrealistic 
$SO(10)$ GUT relations between quark and lepton Yukawa matrices and can 
realize large mixing angles observed in neutrino sector\cite{nomura}. 
Thus, $E_6$ GUT is more interesting because a multiplet includes 
${\bf 16}$ and ${\bf 10}$ as
${\bf 27}={\bf 16}+{\bf 10}+{\bf 1}$\cite{bando}. 
Therefore, $SO(10)$ or $E_6$ GUT is more attractive than $SU(5)$ GUT
in the sense of unification of quark and leptons.
However, in the sense of unification of gauge interactions, $SO(10)$ or
$E_6$ GUT is less natural than $SU(5)$ GUT, because $SO(10)$ or $E_6$ GUT
has additional freedoms which always make the gauge coupling unification
possible in principle. This is because the rank of $SO(10)$
or $E_6$ is larger than that of $SU(5)$ and $G_{SM}$.
But anyway, GUT scenario realizes above two kinds of unification 
simultaneously.

Unfortunately, it is not so easy to obtain the realistic SUSY GUT.
One of the main difficulties is to realize doublet-triplet splitting
in a natural way
\cite{DTsplitting,DW,orbifold}
 with stable proton.
The other difficulty is to avoid unrealistic GUT relations between Yukawa
matrices of quark and leptons, which are related with the unification of
quark and lepton,  and to obtain realistic quark and lepton 
mass matrices.

It is not so easy to solve these problems in a natural way, keeping t he
2 kinds of unifications in GUT scenario.
For example, in the scenario of orbifold GUT which many people have discusseed 
in the literature\cite{orbifold}, 
doublet-triplet splitting is beautifully realized, but
the unification of quark and leptons must be given up even in the sense of
$SU(5)$ to avoid unrealistic GUT relations.

Recently, in a series of papers\cite{maekawa,maekawa2,maekawa3,BM,MY,MY2},
 the interesting GUT scenario with
anomalous $U(1)_A$ gauge symmetry \cite{U(1)}, whose anomaly is cancelled 
by the Green-Schwarz mechanism\cite{GS}, has been proposed with $SO(10)$ unified
group\cite{maekawa} and with $E_6$ unified group\cite{BM,MY}.
One of the biggest difference from the previous works\cite{previous} is that
in our scenario ``generic" interactions (including non-renormalizable 
interactions), namely all the interactions which
are allowed by the symmetry of the theory, are introduced.
Therefore, once we fix the symmetry of the theory 
$(SO(10)$ or $E_6)\times U(1)_A$, we can define the model except $O(1)$
coefficients. The number of input parameters is essentially the number
of the anomalous $U(1)_A$ charges (integer), namely, the number of
multiplets we introduce.
In our scenario, 4 multiplets (
$3\times {\bf 16}+{\bf 10}$) for matter sector and 8 multiplets
($2\times {\bf 45}+2\times ({\bf 16}+{\bf \overline{16}}+{\bf 10})$) for
Higgs sector are
introduced in $SO(10)$ unification, and 3 multiplets
($3\times {\bf 27}$) for matter sector and 8 multiplets
($2\times {\bf 78}+3\times ({\bf 27}+{\bf \overline{27}})$) for Higgs sector
are introduced in $E_6$ unification. 
Here we neglect the number of singlet fields. 
Therefore only 11 or 12 integers are input parameters in our models.

It is amazing that under such a small amount of inputs, following various 
things can be explained in our scenario.
\begin{enumerate}
\item
GUT scales and other symmetry breaking scales are determined by anomalous
$U(1)_A$ charges.
\item
Doublet-triplet splitting is realized in a natural way.
\item
Proton decay via dimension 5 operators is suppressed.
\item
MSSM is obtained at a low energy scale.
\item
Mass spectrum of superheavy fields are fixed by anomalous $U(1)_A$ charges.
\item
Natural gauge coupling unification in a sense of the minimal $SU(5)$ GUT
is realized even if we adopt $SO(10)$ or $E_6$, whose rank is larger than
that of $SU(5)$ or $G_{SM}$, as the unified group.
\item
Realistic quark and lepton masses and mixings are obtained.
\item
$\mu$ problem is solved.
\item
In $E_6$ unification, a condition for suppression of the flavor changing 
neutral current is satisfied.
\end{enumerate}

Predictions of our scenario are
\begin{enumerate}
\item
Bi-large neutrino mixing angles and LMA solution
for the solar neutrino problem;
\item
$U_{e3}\sim O(0.1)$;
\item
CP violation phase in lepton sector $\delta_{CP}\sim O(1)$;
\item
small $\tan\beta\sim 5$;
\item
Smaller unification scale $\Lambda_u<\Lambda_G\sim 2\times 10^{16}$ GeV
(Thus proton decay via dimension 6 operators may be seen in future experiments)
\begin{equation}
\tau(P\rightarrow e\pi)\sim \left\{ 
\begin{array}{cl}
  5\times 10^{33} {\rm years} & (a=-1) \\
  8\times 10^{34} {\rm years} & (a=-1/2)
\end{array} \right. ; 
\end{equation}
\item
Smaller Cutoff scale $\Lambda\sim \Lambda_G<M_{Pl}$
(It may imply the existence of extra dimension in which only gravity
modes can propagate as discussed by Horava-Witten);
\item
Low energy SUSY.
\end{enumerate}

\section{Doublet-triplet splitting}
One of the most interesting feature of anomalous $U(1)_A$ gauge theory is
that the vacuum expectation values (VEV) are determined by anomalous
$U(1)_A$ charges as
\begin{eqnarray}
\VEV{Z^+}&=& 0, \label{VEV+} \\
\VEV{Z^-}&\sim&  \lambda^{-z^-},\label{VEV-}
\end{eqnarray}
where $Z^\pm$ are singlet operators with the charges $z^+>0$ and $z^-<0$, 
and $\lambda=\VEV{\Theta}/\Lambda$. Here $\Theta$ is a Froggatt-Nielsen
field
\cite{FN}.
 Through this paper, we use unit in which the cutoff $\Lambda=1$
and denote all the superfields by uppercase letters and their anomalous
$U(1)_A$ charges by the corresponding lowercase letters.
Such VEVs do not change the order of the coefficients obtained by
the Froggatt-Nielsen mechanism:
\begin{equation}
W=\left(\frac{\Theta}{\Lambda}\right)^{x+y+z}XYZ\rightarrow \lambda^{x+y+z}XYZ,
\end{equation}
if the total charge $x+y+z$ of the operator $XYZ$ is positive.
Using this mechanism, the hierarchy of Yukawa couplings can be realized
\cite{yukawa}
Note that even if the operator $\frac{Z^-}{\Lambda}$ is used instead of 
$\left(\frac{\Theta}{\Lambda}\right)^{-z^-}$ in the interactions, the order
of the coefficients does not change. This feature is critically different
from the naive expectation that the contribution from the higher dimensional 
operators is more suppressed. 
If the total charge $x+y+z$ is negative, such interaction is not allowed by
the anomalous $U(1)_A$ gauge symmetry because only negatively charged fields
have non-vanishing VEVs. This is called SUSY zero mechanism. 
Note that this mechanism leads to the finite number of non-renormalizable 
interactions, and therefore we can control the generic superpotential.

Actually, under the vacua (\ref{VEV+}), the generic 
superpotential to determine the VEVs of $Z^-$ can be written as
\begin{equation}
W=\sum_i^{n_+}W_{Z_i^+}, \label{W}
\end{equation}
where $W_X$ denotes the terms linear in the $X$ field.
This is because the $F$-flatness conditions of negatively charged fields
are automatically satisfied and the terms with more than two positively charged
fields do not contribute in the $F$-flatness condition of positively charged
fields. 

Let us discuss an $SO(10)$ GUT model with anomalous $U(1)_A$ gauge symmetry
in which doublet-triplet splitting is naturally realized. 
The Higgs content is 
listed in Table I. 
\begin{table}
\begin{center}
Table I. Typical values of anomalous $U(1)_A$ charges.\\
\vspace{1mm}
\begin{tabular}{|c|c|c|} 
\hline
                  &   non-vanishing VEV  & vanishing VEV \\
\hline 
{\bf 45}          &   $A(a=-1,-)$        & $A'(a'=3,-)$      \\
{\bf 16}          &   $C(c=-4,+)$        
                  & $C'(c'=3,-)$      \\
${\bf \overline{16}}$&$\bar C(\bar c=-1,+)$ 
                  & $\bar C'(\bar c'=6,-)$ \\
{\bf 10}          &   $H(h=-3,+)$        & $H'(h'=4,-)$      \\
{\bf 1}           &$\Theta(\theta=-1,+)$,$Z(z=-2,-)$,
                  $\bar Z(\bar z=-2,-)$& $S(s=5,+)$ \\
\hline
\end{tabular}
\end{center}
\end{table}
Here the symbols $\pm$ denote the $Z_2$ parity.
The VEVs of the negatively charged Higgs fields are determined by
the superpotential
\begin{equation}
W=W_{A'}+W_{C'}+W_{\bar C'}+W_{H'}+W_S.
\end{equation}
We do not have spaces enough to explain the vacuum structure in detail,
so we here point out only one good feature in realizing the doublet-triplet
splitting. Actually this observation is important in solving doublet-triplet
splitting with generic interactions.

If $-3a\leq a^\prime < -5a$,
the superpotential $W_{A^\prime}$ is in general
written 
\begin{equation}
W_{A^\prime}=\lambda^{a^\prime+a}\alpha A^\prime A+\lambda^{a^\prime+3a}(
\beta(A^\prime A)_{\bf 1}(A^2)_{\bf 1}
+\gamma(A^\prime A)_{\bf 54}(A^2)_{\bf 54}),
\end{equation}
where the suffices {\bf 1} and {\bf 54} indicate the representation 
of the composite
operators under the $SO(10)$ gauge symmetry, and $\alpha$, 
$\beta$ and $\gamma$ are parameters of order 1. Here we assume 
$a+a^\prime+c+\bar c<0$
to forbid the term $\bar C A^\prime A C$, which destabilizes the 
DW form of the VEV $\VEV{A}$. 
The $D$-flatness condition requires the VEV
$\VEV{A}=i\tau_2\times {\rm diag}(x_1,x_2,x_3,x_4,x_5)$, and the 
$F$-flatness conditions of the $A^\prime$ field requires
$x_i(-\alpha\lambda^{-2a}
+(2\beta-\frac{\gamma}{5})(\sum_j x_j^2)+\gamma x_i^2)=0$. 
This allows only two solutions, $x_i^2=0$ and 
$x_i^2=v^2\sim \lambda^{-2a}$. 
Here $N_0=0$ -- 5 is the number of $x_i=0$ solutions.
When $N_0=2$, the vacuum becomes
$\VEV{A({\bf 45})}_{B-L}=\tau_2\times {\rm diag}
(v,v,v,0,0)$, which breaks $SO(10)$ into
$SU(3)_C\times SU(2)_L\times SU(2)_R\times U(1)_{B-L}$
at the scale $\Lambda_A\equiv \VEV{A}\sim \lambda^{-a}$.
 This Dimopoulos-Wilczek form of the VEV plays an
important role in solving the DT splitting problem.
Actually through the interaction
$W=H'AH$, the DW type of the VEV gives superheavy masses only to the 
triplet Higgs, and therefore the doublet Higgs remains massless. 
Taking account of the mass term $H'^2$, only one pair of Higgs doublets
becomes massless. 

Note that the higher terms $A^\prime A^{2L+1}$ $(L>1)$ are 
forbidden by the SUSY zero mechanism. If they were allowed, 
the number of possible VEVs other than the DW form would 
become larger, and thus it would become less natural to obtain 
the DW form. 
This is a critical point of this mechanism, and the anomalous 
$U(1)_A$ gauge symmetry plays an essential role in forbidding 
the undesired terms.

The spinor Higgs fields 
$C$ and $\bar C$ break $SU(2)_R\times U(1)_{B-L}$ into $U(1)_Y$ 
by developing 
$\VEV{C}(=\VEV{\bar C}\equiv \Lambda_C\sim \lambda^{-(c+\bar c)/2}$).
Then this model becomes MSSM at a low energy scale.

\section{Gauge coupling unification}
Unfortunately, the mass spectrum of superheavy Higgs fields does not
respect $SU(5)$ gauge symmetry. Naively thinking, it spoils the
success of gauge coupling unification in SUSY GUT scenario.
In the early stage of these works, we expected that there must be 
tuning parameters to realize the gauge coupling unification because
the rank of $SO(10)$ or $E_6$ is larger than that of $SU(5)$. 
However, the fact is more exciting than what we expected.

In our scenario, the mass spectrum of superheavy fields and GUT breaking
scales are determined by anomalous $U(1)_A$ charges, so the conditions
for gauge coupling unification
\begin{equation}
\alpha_1(\Lambda_u)=\alpha_2(\lambda_u)=\alpha_3(\Lambda_u)
\end{equation}
are rewritten by using anomalous $U(1)_A$ charges, cutoff scale $\Lambda$
and usual GUT scale $\Lambda_G$ as
\begin{eqnarray}
\Lambda\sim \Lambda_G \\
h\sim 0.
\end{eqnarray}
It is miraculous that all the charges except doublet Higgs's are cancelled out.
And the first condition is nothing but for determining the scale of the theory,
and the second condition is essentially the same freedom as that of
colored Higgs mass in the minimal $SU(5)$ GUT because $h$ is anomalous $U(1)_A$
charge not only of doublet Higgs but also of colored Higgs.
Therefore, we have no other parameters to be adjusted than in the minimal
$SU(5)$ GUT. 
Let us explain the situation.
In our scenario, since mass spectrum of superheavy fields and GUT breaking
scales are determined by charges, we can estimate the gauge coupling constants
$\alpha_i(\Lambda_W)$ at a low energy scale from the cutoff scale 
for any charges and any cutoff.
By using the estimated value $\alpha_i(\Lambda_W)$, we can calculate the running
gauge couplings in MSSM. It is surprising that for any charges and for any 
cutoff, three gauge couplings meet at a scale. The scale is the same as the
cutoff scale in our scenario.
If our scenario is real, since it is known that three gauge couplings meet
at the usual GUT scale $\Lambda_G\sim 2\times 10^{16}$ GeV in MSSM, 
the cutoff scale in our scenario must be taken as $\Lambda_G$.
On the other hand, the real unification scale is usually smaller than
the cutoff $\Lambda_u\sim \lambda^{-a}<\Lambda=\Lambda_G$.
Therefore proton decay via dimension 6 operators is interesting in
our scenario, and actually, rough estimation leads to
\begin{equation}
\tau(P\rightarrow e\pi)\sim \left\{ 
\begin{array}{cl}
  5\times 10^{33} {\rm years} & (a=-1) \\
  8\times 10^{34} {\rm years} & (a=-1/2)
\end{array} \right. .
\end{equation}
We would like to emphasize that the above picture is independent of the 
details of Higgs sector, namely independent
of their charges and of how to realize doublet-triplet splitting.
Actually sufficient conditions for the miracle cancellation\cite{MY2} are
\begin{enumerate}
\item
The unification group is simple.
\item
\begin{equation}
\VEV{O_i}\sim \left\{ 
\begin{array}{ccl}
  \lambda^{-o_i} & \quad & o_i\leq 0 \\
  0              & \quad & o_i>0
\end{array} \right. . 
\end{equation}
\item
At a low energy scale, MSSM(+singlets) is realized.
\end{enumerate}
The vacuum structure is naturally obtained if generic interactions are 
introduced and $F$-flatness conditions determine the scale of the VEVs.
Thus the above picture is fairly general one in GUT with anomalous $U(1)_A$
gauge symmetry.

\section{Quark and lepton mass matrices}
One of the most attractive features of grand unified theory is to 
unify the quark and lepton into fewer multiplets. For example, 
in $SO(10)$ GUT scenario, a ${\bf 16}$ representation field contains
one family quark and lepton fields including right-handed neutrino
field. However, this attractive feature directly leads to unrealistic
Yukawa relations. For example, if we introduce $3\times{\bf 16}$ 
$\Psi_i (i=1,2,3)$ for 3 family quark and leptons, the Yukawa couplings
are obtained from the interaction
\begin{equation}
W=Y_{ij}\Psi_i\Psi_j H
\end{equation}
as $Y_u=Y_d$ and $Y_d=Y_e,$
which lead to unrealistic mass relation.
We have to pick up the VEV $\VEV{C}$ in the Yukawa matrices to avoid
the former unrealistic relation $Y_u=Y_d$, and the VEV $\VEV{A}$
to avoid the latter unrealistic relation $Y_d=Y_e$.
In our scenario, we introduce an additional matter field $T({\bf 10})$.
Then after breaking the GUT gauge group into the standard model gauge
group, one pair of vector-like fields ${\bf 5}$ and ${\bf \bar 5}$ of
$SU(5)$ becomes massive. The mass matrix is 
obtained from the interaction
\begin{equation}
W=\lambda^{\psi_i+t+c}\Psi_iTC+\lambda^{2t}T^2
\end{equation}
as 
\begin{equation}
{\bf 5}_T ( \lambda^{t+\psi_1+(c-\bar c)/2}, 
               \lambda^{t+\psi_2+(c-\bar c)/2}, 
               \lambda^{t+\psi_3+(c-\bar c)/2}, \lambda^{2t})
\left(
\begin{array}{c}  {\bf \bar 5}_{\Psi1} \\ {\bf \bar 5}_{\Psi2} \\ 
{\bf \bar 5}_{\Psi3}
 \\ {\bf \bar 5_T}
\end{array}
\right),
\end{equation}
where actually the VEV $\VEV{\bar C}=\VEV{C}\sim \lambda^{-(c+\bar c)/2}$
appear in the mass matrix.
Since $\psi_3<\psi_2<\psi_1$, 
the massive mode ${\bf \bar 5}_M$, the partner of ${\bf 5}_T$, must
be either 
${\bf \bar 5}_{\Psi3}(\Delta\equiv 2t-(t+\psi_3+(c-\bar c)/2)>0)$ or
${\bf \bar 5}_T(\Delta<0)$. 
The former case is interesting, and in this case,
the three massless modes 
$({\bf \bar 5}_1, {\bf \bar 5}_2, {\bf \bar 5}_3) $ 
can be written 
$({\bf \bar 5}_{\Psi1}+\lambda^{\psi_1-\psi_3}{\bf \bar 5}_{\Psi3}, 
{\bf \bar 5}_T+ \lambda^{\Delta}{\bf \bar 5}_{\Psi3},
{\bf \bar 5}_{\Psi2}+\lambda^{\psi_2-\psi_3}{\bf \bar 5}_{\Psi3})$. 
If we adopt their charges $(\psi_1,\psi_2,\psi_3,t)=(9/2,7/2,3/2,5/2)$
in addition to the charges of Higgs fields, then we can estimate quark 
and lepton mass matrices as
\begin{equation}
M_u=\left(
\begin{array}{ccc}
\lambda^6 & \lambda^5 & \lambda^3 \\
\lambda^5 & \lambda^4 & \lambda^2 \\
\lambda^3 & \lambda^2 & 1  
\end{array}
\right)\VEV{H_u},\quad
M_d=M_e=\lambda^2\left(
\begin{array}{ccc}
\lambda^4 & \lambda^{3.5} & \lambda^3 \\
\lambda^3 & \lambda^{2.5} & \lambda^2 \\
\lambda^1 & \lambda^{0.5} & 1
\end{array}
\right)\VEV{H_d}.
\label{quark}
\end{equation}
And for the neutrino sector, we take into account the interaction
\begin{equation}
\lambda^{\psi_i+\psi_j+2\bar c}\Psi_i\Psi_j\bar C\bar C,
\end{equation}
which lead to the right-handed neutrino masses
\begin{equation}
M_R=\lambda^{\psi_i+\psi_j+2\bar c}\VEV{\bar C}^2
=\lambda^{2n+\bar c-c}\left(
\begin{array}{ccc}
\lambda^6 & \lambda^5 & \lambda^3 \\
\lambda^5 & \lambda^4 & \lambda^2 \\
\lambda^3   & \lambda^2         & 1
\end{array}
\right).
\end{equation}
Since the Dirac neutrino mass is given by
\begin{equation}
M_{\nu_D}=\lambda^2\left(
\begin{array}{ccc}
\lambda^4 & \lambda^3 & \lambda \\
\lambda^{3.5} & \lambda^{2.5} & \lambda^{0.5} \\
\lambda^3   & \lambda^2         & 1 
\end{array}
\right)\VEV{H_u}\eta,
\end{equation}
the neutrino mass matrix is obtained by the seesaw mechanism
as
\begin{equation}
M_\nu=M_{\nu_D}M_R^{-1}M_{\nu_D}^T=\lambda^{4-2n+c-\bar c}\left(
\begin{array}{ccc}
\lambda^2          & \lambda^{1.5}  & \lambda \\
\lambda^{1.5} & \lambda & \lambda^{0.5} \\
\lambda            & \lambda^{0.5}  & 1 
\end{array}
\right)\VEV{H_u}^2\eta^2.
\end{equation}
Note that the ratio $\frac{m_{\nu_\mu}}{m_{\nu_\tau}}\sim \lambda$
is realized, that predicts LMA solution for the solar neutrino problem. 
It is interesting that we obtain the small mixing angles for
the Cabibbo-Kobayashi-Maskawa matrix
\begin{equation}
U_{\rm CKM}=
\left(
\begin{array}{ccc}
1 & \lambda &  \lambda^3 \\
\lambda & 1 & \lambda^2 \\
\lambda^3 & \lambda^2 & 1
\end{array}
\right),
\label{CKM}
\end{equation}
and the large mixing angles for the Maki-Nakagawa-Sakata matrix
\begin{equation}
U_{\rm MNS}=
\left(
\begin{array}{ccc}
1 & \lambda^{0.5} &  \lambda \\
\lambda^{0.5} & 1 & \lambda^{0.5} \\
\lambda & \lambda^{0.5} & 1
\end{array}
\right). 
\end{equation}
Since we use a rule that 
$\lambda^{0.5}+\lambda^{0.5}\sim \lambda^{0.5}$ in calculating the MNS
matrix and $\lambda^{0.5}\sim 0.5$, this model gives large mixing angles
\cite{SK,SNO}
for the atmospheric neutrino problem and for the solar neutrino problem.
And $U_{e3}\sim \lambda$ is predicted, which is around the present upper
limit given by CHOOZ\cite{CHOOZ}.
At this stage, the unrealistic GUT relation $Y_d=Y_e$ still remains.
However, in our scenario, the same amount of the Yukawa couplings are
given by the higher dimensional interactions
\begin{equation}
W=\lambda^{\psi_i+\psi_j+n a+h}\Psi_iA^n\Psi_jH
\end{equation}
by developing the VEV $\VEV{A}\sim \lambda^{-a}$.
It is critical that the Yukawa couplings from the higher dimensional 
interactions have not kept the unrealistic GUT relation.
Usually, the corrections from such higher dimensional interactions are
suppressed by the factor $\frac{\VEV{A}}{\Lambda}$. But in our scenario,
the suppression factor $\frac{\VEV{A}}{\Lambda}$ is just cancelled by the
enhancement factor $\lambda^{a}$ in the coefficients, and therefore
we can obtain the same order coefficients as from the tree interaction.
This is an attractive feature in our scenario, and 
the realistic mass matrices are naturally obtained.

\section{Discussions and summary}
It is familiar to introduce generic interactions to explain
hierarchical structure of Yukawa couplings, using Froggatt-Nielsen
mechanism. However, it has not been done to introduce generic
interactions in Higgs sector. This is because if we introduce generic
interactions in Higgs sector, the gauge group is expected to be broken
maximally into $U(1)^n$. In this talk, we showed that even with generic
interactions, non-abelian structure can be obtained and doublet-triplet
splitting is realized. If we adopt $SO(10)$ or $E_6$ gauge group to
unify quark and lepton, usually it becomes less natural to realize
coupling unification. However, in our scenario, natural gauge coupling
unification is realized in $SO(10)$ or $E_6$ GUT.
Moreover, we obtained realistic quark and lepton mass matrices.

The proton decay via dimension 5 operators\cite{SY} is suppressed
in our scenario, because effective colored Higgs mass is given by
$m_H^c\sim \lambda^{2h}$ and the dimension 5 operators are suppressed
due to anomalous $U(1)_A$ gauge symmetry. On the other hand, proton decay
via dimension 6 operators is reachable, because the cutoff scale must be
around the usual GUT scale $\Lambda_G\sim 2\times 10^{16}$ GeV and the 
unification scale is given by $\lambda^{-a}$. If we adopt $a=-1$, then
the lifetime of the proton is roughly estimated as
\begin{equation}
\tau_p(p\rightarrow e\pi^0)\sim 5\times 10^{33}
\left(\frac{\Lambda_A}{5\times 10^{15}\ {\rm GeV}}\right)^4
\left(\frac{0.015({\rm GeV})^3}{\alpha}\right)^2  {\rm years},
\end{equation}
which is near the present experimental lower bound\cite{SKproton}.
Here $\alpha$ is the hadron matrix element parameter. Here we use the
formula\cite{Murayama} and the value $\alpha$ given by lattice calculation
\cite{lattice}.
Though the prediction is strongly dependent on the actual unification
scale which is dependent on the order one coefficients, this rough estimation
gives a strong motivation for the future experiments for proton decay search.


\begin{thebibliography}{99}
\bibitem{maekawa} N. Maekawa, {\it Prog. Theor. Phys.} {\bf 106}, 401
                  (2001); hep-ph/0110276.
\bibitem{maekawa2} N. Maekawa, {\it Phys. Lett.} {\bf B521}, 42 (2001). 
\bibitem{maekawa3} N. Maekawa, {\it Prog. Theor. Phys.} {\bf 107}, 597 (2002). 
\bibitem{BM}       M. Bando and M. Maekawa, {\it Prog. Theor. Phys.} {\bf 106},
                   1255 (2001).
\bibitem{MY}      N. Maekawa and T. Yamashita, {\it Prog. Theor. Phys.} 
                  {\bf 107}, 1201 (2002). 
\bibitem{georgi}  H. Georgi and S.L. Glashow, {\it Phys. Rev. Lett.}
                  {\bf 32}, 438 (1974).
\bibitem{nomura}  Y. Nomura and T. Yanagida, {\it Phys. Rev.} {\bf D59}, 017303
                  (1999).
\bibitem{bando}  M. Bando and T. Kugo, {\it Prog. Theor. Phys.} {\bf 101},
                 1313 (1999); \\
                 M. Bando, T. Kugo and K. Yoshioka, {\it Prog. Theor. Phys.}
                 {\bf 104}, 211 (2000).
\bibitem{DTsplitting} E. Witten, {\it Phys. Lett.} {\bf B105}, 267 (1981);\\
                      A. Masiero, D.V. Nanopoulos, K. Tamvakis and T.Yanagida,
                      {\it Phys. Lett.} {\bf 115}, 380 (1982);\\
                      B. Grinstein, {\it Nucl. Phys.} {\bf B206}, 387 (1982);\\
                      K. Inoue, A. Kakuto and T. Takano,
                      {\it Prog. Theor. Phys.} {\bf 75}, 664 (1986);\\
                      E. Witten, {\it Nucl. Phys.} {\bf B258},75 (1985);\\
                      T. Yanagida, {\it Phys. Lett.} {\bf B344}, 211 (1995);\\
\bibitem{DW}       S. Dimopoulos and F. Wilczek, NSF-ITP-82-07;\\
                   M. Srednicki, {\it Nucl. Phys.} {\bf B202}, 327 (1982);\\
                   S.M. Barr and S. Raby, {\it Phys. Rev. Lett.}
                   {\bf 79}, 4748 (1997).
\bibitem{orbifold}   Y. Kawamura, {\it Prog. Theor. Phys.} {\bf 105}, 
                      691 (2001);{\it ibid} {\bf 105}, 999 (2001);\\
                   L. Hall and Y. Nomura, {\it Phys. Rev.} {\bf D64},
                  055003 (2001).
\bibitem{MY2}     N. Maekawa and T. Yamashita, {\it Prog. Theor. Phys.} 
                  {\bf 108}, 719 (2002); hep-ph/0209217.
\bibitem{U(1)}    E.~Witten,  {\it Phys. Lett.} {\bf B149}, 351 (1984);\\
                  M.~Dine, N.~Seiberg and E.~Witten,
                  {\it Nucl. Phys.} {\bf B289}, 589 (1987);\\
                  J.J.~Atick, L.J.~Dixon and A.~Sen,
                  {\it Nucl. Phys.} {\bf B292}, 109 (1987);\\
                  M.~Dine, I.~Ichinose and N.~Seiberg,
                  {\it Nucl. Phys.}  {\bf B293}, 253 (1987).
\bibitem{GS}      M.~Green and J.~Schwarz,
                  {\it Phys. Lett.} {\bf B149}, 117 (1984).
\bibitem{previous}   L.J. Hall and S. Raby, {\it Phys. Rev.} {\bf D51}, 
                  6524 (1995);
                  G. Dvali and S. Pokorski, {\it Phys. Rev. Lett.} {\bf 78},
                  807 (1997); Z. Berezhiani and Z. Tavartkiladze, 
                  {\it Phys. Lett.}
                  {\bf B396} 150 (1997);{\it ibid} {\bf B409}, 220 (1997);
                  G. Dvali and A. Riotto, {\it Phys. Lett.} {\bf B417}, 20 
                  (1998);
                  K.-I. Izawa, K. Kurosawa, Y.Nomura and T.Yanagida,
                     {\it Phys. Rev.} {\bf D60}, 115016 (1999);
                  Q. Shafi and Z. Tavartkiladze, {\it Phys. Lett.} {\bf B459}, 
                  563
                  (1999); {\it ibid} {\bf B482}, 145 (2000); {\it ibid} 
                  {\bf B487}, 145 (2000); {\it Nucl. Phys.} {\bf B573}, 40 
                  (2000);
                  J.L. Chkareuli, C.D. Froggatt, I.G. Gogoladze 
                  and A.B. Kobakhidze, {\it Nucl. Phys.} {\bf B594}, 23 (2001).
\bibitem{FN}      C.D. Froggatt and H.B. Nielsen,
                  {\it Nucl. Phys.} {\bf B147}, 277 (1979).
\bibitem{yukawa}  L. Ib\'a\~nez and G.G. Ross,
                  {\it Phys. Lett.} {\bf B332}, 100 (1994);
                  P. Bin\'etruy and P. Ramond,
                  {\it Phys. Lett.} {\bf B350}, 49 (1995);
                  E. Dudas, S. Pokorski and C.A. Savoy,
                  {\it Phys. Lett.} {\bf B356}, 45 (1995);
                  P. Bin\'etruy, S. Lavignac and P. Ramond,
                  {\it Nucl. Phys.} {\bf B477}, 353 (1996); 
                   P. Bin\'etruy, S. Lavignac, S. Petcov and P. Ramond,
                  {\it Nucl. Phys.} {\bf B496}, 3 (1997);
                  H. Dreiner, G.K. Leontaris, S. Lola, G.G. Ross and C. Scheich,
                  {\it Nucl. Phys.} {\bf B436}, 461 (1995);
                   C.H. Albright and S. Nandi, Mod. Phys. Lett. {\bf A11}, 737
                   (1996); Phys. Rev. {\bf D53}, 2699 (1996); 
                   Y. Nomura and T. Sugimoto, Phys. Rev. {\bf D61}, 093003 
                   (2000).
\bibitem{SK}        Y. Fukuda et al.(The Super-Kamiokande Collaboration),
                    {\it Phys. Lett.} {\bf B436}, 33 (1998);
                    {\it Phys. Rev. Lett.} {\bf 81}, 1562 (1998);
                    {\it Phys. Rev. Lett.} {\bf 86}, 5656 (2001).
\bibitem{SNO}      The SNO Collaboration, {\it Phys. Rev. Lett.} {\bf 89},
                   011301 (2002); {\it ibid} {\bf 89}, 011302.
\bibitem{CHOOZ}   The CHOOZ Collaboration, {\it Phys. Lett.} {\bf
                  B420}, 397 (1998).
\bibitem{SY}     N. Sakai and T. Yanagida, {\it Nucl. Phys.} {\bf
                  B197}, 533 (1982).
\bibitem{SKproton} Super-Kamiokande Collaboration, {\it Phys. Rev. Lett.}
                  {\bf 81}, 3319 (1998); {\it ibid} {\bf 83}, 1529 (1999).
\bibitem{Murayama} H. Murayama and A. Pierce, Phys. Rev. D {\bf65}, 
                     055009 (2002).
\bibitem{lattice} JLQCD Collaboration, S. Aoki et al., {\it Phys. Rev.}
                  {\bf D62}, 014506 (2000).
\end{thebibliography}
\end{document}